\newcommand{\CLASSINPUTtoptextmargin}{.72in} % 0.705inc is top required for globecom  edas %.71 (6pages)
\newcommand{\CLASSINPUTbottomtextmargin}{.925in} %0.326in is OK for v04 %.335 (6pages)
\newcommand{\CLASSINPUTinnersidemargin}{.575in} %.575in is boundary for globecom  edas %.585 was ICC number
\newcommand{\CLASSINPUToutersidemargin}{0.68in} %0.68in is boundary for globecom  edas
\documentclass[conference,A4paper]{IEEEtran}
\IEEEoverridecommandlockouts
\renewcommand{\baselinestretch}{0.92}

\usepackage{textcomp}
\usepackage{xcolor}
\usepackage{booktabs} % For formal tables
\usepackage[linesnumbered,ruled]{algorithm2e}
\newcommand{\images}{./images}
\usepackage{mathrsfs}
\usepackage{amsmath}

\usepackage{graphicx}
\usepackage{epstopdf}
\usepackage[caption=true,font=footnotesize]{subfig}
\usepackage[font={footnotesize}]{caption}
\usepackage{cite}
\renewcommand{\IEEEQED}{\IEEEQEDopen}

\begin{document}
%\endlinechar=-1\relax

\title{Coordinated Container Migration and \\Base Station Handover in Mobile Edge Computing
}

\author{
\IEEEauthorblockN{%
	Mao~V.~Ngo\IEEEauthorrefmark{1},%\IEEEauthorrefmark{2},
    Tie~Luo\IEEEauthorrefmark{2},
	Hieu~T.~Hoang\IEEEauthorrefmark{1}, and
	Tony~Q.S.~Quek\IEEEauthorrefmark{1}}%
\IEEEauthorblockA{\IEEEauthorrefmark{1}%
	Singapore University of Technology and Design,
	Singapore}%
%\IEEEauthorblockA{\IEEEauthorrefmark{2}%
%	Institute for Infocomm Research,
%	A*STAR,
%	Singapore}%
\IEEEauthorblockA{\IEEEauthorrefmark{2}%
    Department of Computer Science,
    Missouri University of Science and Technology,
    USA}%
\IEEEauthorblockA{%
	\texttt{vanmao\_ngo@sutd.edu.sg},
    \texttt{tluo@mst.edu},
	\texttt{hthhieu@gmail.com},
	\texttt{tonyquek@sutd.edu.sg}}%
%\thanks{%
%	\hrule width 0.33\columnwidth \vskip5pt
%	The work of xx was supported }%
} % author

\maketitle

\begin{abstract}
%Migration in edge computing system is an important requirement to provide smooth experience, non-interruption for stringent low latency emerging applications to mobile users (MUs).
Offloading computationally intensive tasks from mobile users (MUs) to a virtualized environment such as containers on a nearby edge server, can significantly reduce processing time and hence end-to-end (E2E) delay.
However, when users are mobile, such containers need to be {\em migrated} to other edge servers located closer to the MUs to keep the E2E delay low. %Indeed, {\em container migration} between edge servers is an indispensable feature of mobile edge computing.
Meanwhile, the mobility of MUs necessitates {\em handover} among base stations in order to keep the wireless connections between MUs and base stations uninterrupted.
In this paper, we address the joint problem of container migration and base-station handover by proposing a coordinated migration-handover mechanism, with the objective of achieving low E2E delay and minimizing service interruption.
The mechanism determines the optimal destinations and time for migration and handover in a coordinated manner, along with a delta checkpoint technique that we propose.
%Our mechanism keeps track of the movement of MUs and the resource utilization of edge servers, and initiates container migration at the best time in coordination with base-station handover. It minimizes service downtime using a , and the benefit is amplified when performed over wide area networks (WANs).
We implement a testbed edge computing system with our proposed coordinated migration-handover mechanism, and evaluate the performance using real-world applications implemented with Docker container (an industry-standard).
%The results demonstrate that the mechanism achieves the lowest service downtime and E2E delay compared to other mechanisms that we also implement and evaluate.
The results demonstrate that our mechanism achieves 30\%-40\% lower service downtime and 13\%-22\% lower E2E delay as compared to other mechanisms. % that we also implement and evaluate.
Our work is instrumental in offering smooth user experience in mobile edge computing.
\end{abstract}

%\begin{IEEEkeywords}Docker, container, offloaded service, fog computing, mobile computing, handoff, low latency\end{IEEEkeywords}

% Start
\section{Introduction}
\iffalse
Mobile devices are ubiquitous in our lives, and are hosting more and more mobile applications that demand increasingly more computational resources.
While hardware advancement of mobile devices has never ceased, the gap between supply and demand in terms of computational resources still exists.
To this end, cloud computing enables mobile users (MUs) to offload computationally intensive tasks to remote, powerful cloud servers.
However, the delay incurred by such offloading with clouds do not satisfy stringent latency requirements of emerging applications, such as %virtual reality (VR), augmented reality (AR),
smart collision avoidance in transportation, and surveillance systems using face/object recognition.
\fi

As the next evolution of computing paradigm, mobile edge computing (MEC) brings computation, storage, and communication much closer to end-users \cite{Chen_SEC2017, 5GCORALD2}.
It allows a mobile user (MU) to offload its computationally intensive tasks (e.g., image processing) to nearby {\em edge servers} to significantly reduce the end-to-end (E2E) delay as compared to using the cloud counterpart \cite{Chen_SEC2017}.
An edge server hosts multiple {\em containers} or {\em offloaded services} (the latter is just the former in the running status, so we will use them interchangeably depending on the context), each of which runs a task offloaded by a user.
While a virtual machine (VM) creates a full guest operating system (OS), containers are a lightweight virtualization technology that shares the same OS kernel and isolates the application processes from the rest of the system.
Therefore, containers not only solve the environment-dependence issue but also notably reduce memory footprint, initialization, and migration overhead \cite{MachenIEEE2018}.
%Containers are a lightweight virtualization technology as compared to virtual machines (VM), due to its notable reduction of memory footprint and migration overhead \cite{MachenIEEE2018}. %initialization
%For example, Z. Chen \textit{et al.} \cite{Chen_SEC2017} study latency of some edge computing applications with a Google-glass-like device as wearable cognitive assistance to enhance some tasks required timely response such as Lego, drawing, or playing ping-pong.
% since it reduces significant latency from the MU to the service's location \cite{Chen_SEC2017}.

However, when a user who is served by a \textit{stateful} offloaded service\footnote{A stateful service (such as a video game) maintains its state information of users context \cite{ETSI_GS_MEC_021} for future sessions.} moves, and if its associated edge server does not change, the latency advantage will start to degenerate toward the original cloud-based offloading and can become even worse \cite{Chen_SEC2017}.
Therefore, it is necessary to perform {\em container migration} \cite{Mirkin08containerscheckpointing} from the current edge server to another edge server that is closer to the MU, with minimal service interruption (i.e., downtime).
Since the moving trajectories of MUs are typically unknown a priori, it is challenging to know the best time and destination edge server to migrate the current container.
Furthermore, it is also desirable to be able to migrate over wide area networks (WANs) rather than LANs only \cite{Chen_SEC2017, Lele17}, which constitutes another challenge.

Besides container migration, the wireless connection between a MU and its associated base station (BS)---cellular BS or WiFi access point---needs to be handed over to another BS as well (note that a BS may or may not be collocated with an edge server). Although {\em handover} has been well studied in cellular networks \cite{Rappaport01book}, the key difference here is that, in MEC, the handover between BSs takes place in conjunction with container migration between edge servers, while in cellular networks, all computation tasks are hosted in a central server and hence there is no need for migration.
Therefore, the handover process and the migration process need to be coordinated with optimized timing and destination hosts in order to achieve minimal service downtime and provide smooth user experience.
%Therefore, the handover and migration processes need to be coordinated with optimized timing and destination hosts in order to minimize service downtime and provide smooth user experience.
%The key thing to reduce the service downtime is the migration process needs to be started before the MU moves to another BS, otherwise the MU suffers long service downtime and high E2E delay because it is still served by the old edge server.
%So that, this paper addresses the problem of handle collaboratively handover wireless connection together with migration offloaded service to meet MUs' demands.

%To the best of our knowledge, this paper is the first work that addresses the challenge of joint migration and handover in MEC to jointly minimize service downtime and low E2E delay.
Our main contributions are as follows:
\begin{itemize}
	\itemsep0em
    \item We present a MEC architecture that assists container migration and BS handover (as well as monitoring and deployment) for MUs with user context transfer (Section~\ref{sec:systemArchitecture}).
    %The proposed edge computing system makes a migration-handover plan based on the edge computing resource and the MU's connection status.

    \item We propose a \textit{coordinated migration-handover mechanism} to minimize E2E delay and service interruption. The mechanism consists of two parts:
    \iffalse
     \begin{itemize}
     	\itemsep0em
     	\item Part A: \textit{Optimal Placement} which determines the best destination edge servers for migration and the best BSs for handover (Section~\ref{sec:handoverMigrationPlanner}).
     	\item Part B: \textit{Best Triggering Time} which determines the coordinated time to trigger various stages of the migration process (based on our proposed \textit{delta checkpoint} technique) and the handover process (Section~\ref{sec:orchestratedMechanism}).
     \end{itemize}
    \else
    (i) \textit{Optimal Placement} which determines the best destination edge servers for migration and the best BSs for handover (Section~\ref{sec:handoverMigrationPlanner}); and
    (ii) \textit{Best Triggering Time} which determines the coordinated time to trigger various stages of the migration process (based on our proposed \textit{delta checkpoint} technique) and the handover process (Section~\ref{sec:orchestratedMechanism}).
    \fi
    %\item We formulate an optimization problem to determine destinations for service offloading with consideration of both handover and migration constraints, in order to achieve low E2E delay and minimize service interruption (Section~\ref{sec:handoverMigrationPlanner}).

    %\item We propose a mechanism to orchestrate container migration and handover processes (Section~\ref{sec:orchestratedMechanism}).
     % based on the estimation of migration time, time to handover, and total service downtime. (Section~\ref{sec:orchestratedMechanism}).

    \item We built a MEC testbed, implemented our proposed coordinated migration-handover mechanism, and evaluated its performance using real-world applications that we developed.
    The experimental results show that the proposed mechanism outperforms one baseline and two state-of-the-art mechanisms~(Section~\ref{sec:experimentalSetup}). The open-source code is available at
    %\url{https://gitlab.com/ngovanmao/edgecomputing}.
     \textit{https://gitlab.com/ngovanmao/edgecomputing}.
\end{itemize}

%----------------------------------------------
\section{Related Work}
\label{sec:background}
%----------------------------------------------
VM migration on edge computing system has been considered in several works \cite{Bittencourt2015, MachenIEEE2018, ZhuVTC2018, XiangSunICC2016,5GCORALD2, Lele17, Nasrin_ICC2018, Bellavista_IWCMC2017}.
However, most of the systems \cite{Bittencourt2015,MachenIEEE2018,XiangSunICC2016, Nasrin_ICC2018} focus on VM migration over LANs which is more relevant to cloud computing; some \cite{Bellavista_IWCMC2017, ZhuVTC2018, 5GCORALD2} consider computation task offloading for mobile users but have overlooked the network condition of mobile devices.
Nasrin \textit{et al.}~\cite{Nasrin_ICC2018} proposed a SharedMEC system to reduce unnecessary migrations during handover among femtocell BSs.
In our work, we consider a comprehensive system including both placement and migration of \textit{stateful} offloaded services (i.e., containers), together with handover of wireless connections for mobile users.
Prior work evaluates the performance of migration using simulations \cite{XiangSunICC2016, Bittencourt2015, Nasrin_ICC2018}, whereas we evaluate both migration and handover using an actual MEC testbed with real applications deployed as Docker containers.
In addition, some papers describe MEC architectures but without migration for MUs, whereas we present an architecture that fills this gap.
Finally, we jointly minimize E2E delay and total service downtime by taking both theoretical and practical approaches.

\section{Edge Computing System Architecture}
\label{sec:systemArchitecture}
%----------------------------------------------

\begin{figure*}[t]
	\centering
	\includegraphics[width=0.90\linewidth]{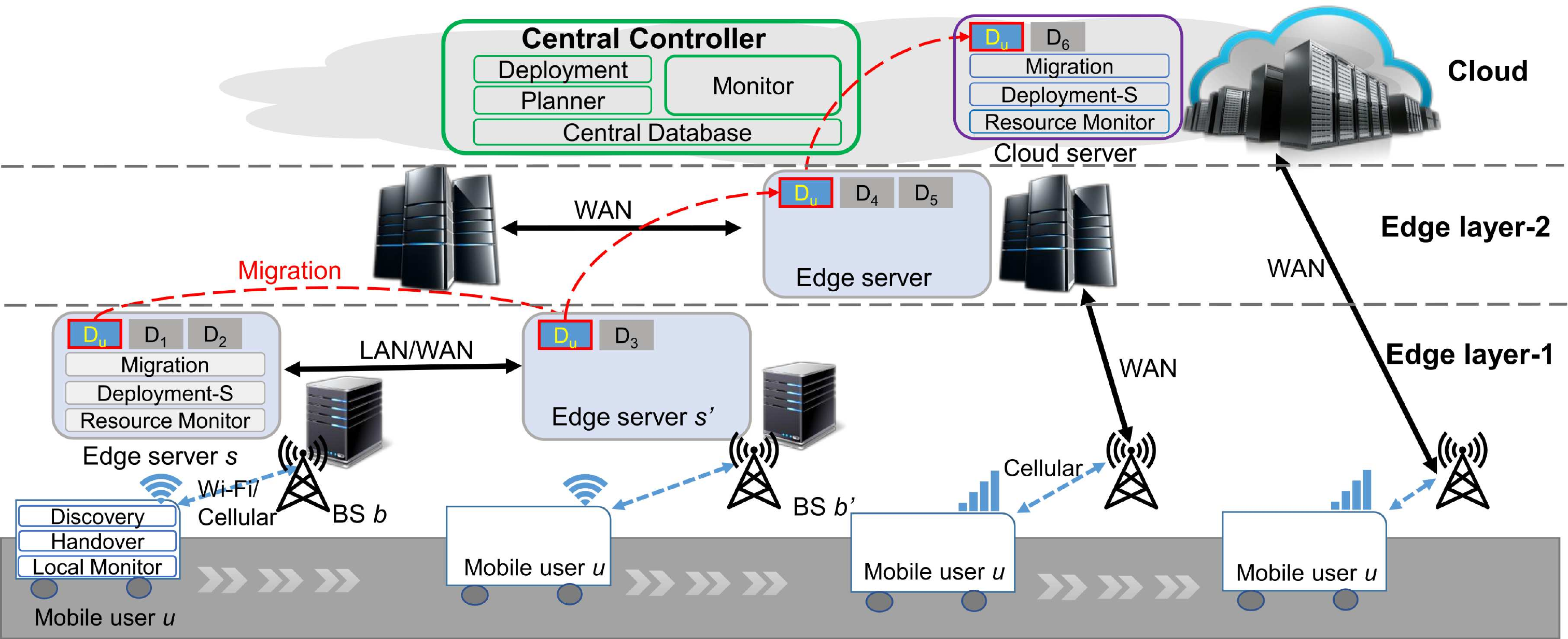}
	\caption{Edge computing system architecture that supports both container migration and base station handover.}
	\label{fig:SystemArch}
\end{figure*}

%We present a MEC system architecture as shown in Fig. \ref{fig:SystemArch}. Its hierarchical structure includes three layers \cite{5GCORALD2, Bellavista_IWCMC2017}. %CAMERA:\cite{5GCORALD2, MaoDCN2018}.
We present a hierarchical MEC system including three layers as shown in Fig.~\ref{fig:SystemArch}.
The layer-1 edge servers are collocated with BSs at the access network. The layer-2 servers are further away from MUs but have more computational power; they are fewer and connect to sparsely distributed BSs (such as in rural areas) which are not collocated with edge servers. Cloud resides at the third and highest layer, and hosts one or more (cloud) servers or VMs that have the same internal modules as the edge servers.
%The edge computing system has three key components: a central controller, edge servers, and mobile users.

%\textbf{Central controller} can help MUs allocate offloaded services by placement service, monitor service, and planner module with centralized database.
The cloud hosts a single \textbf{central controller} which maintains a global view of the whole system. It gathers information from all the edge servers, BSs, and MUs, and stores it in a \textit{Central Database}. The \textit{Deployment} module listens to and receives offloading requests from MUs, invokes a {\em Planner} module to find an edge server for placing offloaded services, and issues this instruction to the corresponding edge server. The \textit{Monitor} module monitors MU-related performance such as E2E delay and received signal strength indication (RSSI).
If the performance fails to meet a pre-defined service level agreement (SLA), or the MU needs a handover, it triggers the {\em Planner} module to find a better placement scheme.
The \textit{Planner} module combines all the monitored information to make a \textit{migration-handover plan} which includes:
{placement of edge server (where to host the container),
placement of BS	(which BS to connect to the MU),
the time of migration, and
the time of handover}.

An \textbf{edge server} uses a \textit{Deployment-S} module to deploy the container based on instructions issued by the central controller, notify ready-to-use to MUs, and update its status to the central controller.
The \textit{Resource Monitor} module monitors the resource utilization with respect to computation, memory, storage, and network of the edge server and its containers.
The \textit{Migration} module on the source and destination edge servers collaborate with each other to migrate containers based on instructions from the central controller.
% from one server to another which consists of \textit{source-migration} and \textit{dest-migration}.
%The source-migration executes the pre-migration, migration based on instructions from the central controller.
%The dest-migration cooperates with the source-migration to restore the container, and notifies to the MUs and the central controller.

\textbf{Mobile users} use a \textit{Discovery} module to request to the MEC system for offloading tasks.
The \textit{Local Monitor} module regularly sends E2E delay and RSSI with respect to nearby BSs to the central controller.

Our MEC system design follows the ETSI specifications on application mobility service \cite{ETSI_GS_MEC_021} with the implementation approach of \textit{MEC assisted user context transfer}.
%\footnote{
%\red{We used the publish-subscribe JSON-based messages for communicating control signal instead of the HTTP method as the ETSI specification \cite{ETSI_GS_MEC_021}.}
%Instead of using HTTP for communicating control signal suggested by the ETSI specification \cite{ETSI_GS_MEC_021}, we used the publish-subscribe JSON-based messages for flexibility and scalability.
%}

%----------------------------------------------
%\section{Mechanism Part A: Optimal Placement}
\section{Design of Coordinated Mechanism}
\label{sec:DesignCoordinatedMechanism}
%----------------------------------------------

%\subsection{Problem Formulation}
\subsection{Optimal Placement}
\label{sec:handoverMigrationPlanner}

This subsection deals with determining the optimal destinations---edge servers and BSs---for migration and handover. The objective is to maintain a low E2E delay for MUs with minimal service interruption.

Consider a MEC system that consists of a set of MUs $\mathcal{U}$ %(e.g., autonomous vehicles, robots, AR/VR gaming terminals),
a set of BSs $\mathcal{B}$, and a set of servers $\mathcal{S}$.
%A server $s\in\mathcal S$ can be an edge server or a cloud server, which are the same with just different resource capacities.
We let $\mathcal{S}$ include the cloud server as well because it is the same as an edge server except for different resource capacities.
At time $t$, a MU $u\in\mathcal U$ who is connected to a BS $b$ continuously offloads tasks to an offloaded service deployed on a server $s$ as a container $D_u$.
Denote by $z^{t}_{ubs} \in \{1, 0 \}$ whether or not $u$ is connected to the BS $b$, and its offloaded service is allocated to the server $s$, at $t$.
%Denote by $x_{us}^{t}$, a binary variable, whether or not the offloaded service of the MU $u$ is allocated on the server $s$ at $t$. Hence for the above case, $x_{us}^{t}=1$.
%Similarly, denote by $y_{ub}^t$, a binary variable, whether or not the MU $u$ is associated with a BS $b$ at $t$. Hence for the above, $y_{ub}^{t}=1$.
%Hence for the above case, $x_{us}^{t}=1$ and $y_{ub}^{t}=1$.
Denote by $d_{ubs}^{t}$ the E2E delay which is defined as the interval between time $t$ when $u$ sends an offloading task to a server $s$ via its associated BS $b$, and the time when $u$ receives a task execution result from some server via some BS.

Suppose at a future time $t'=t+\Delta t$, the MU moves from the current BS $b$ to the vicinity of another BS $b'$.
We need to find a new server $s'$ to host the container and a new BS $b'$ to keep the wireless connection for the MU.
In short, we need to find a new variable $z_{ub's'}^{t'}$ at $t'$.
%Define a new variable $z_{ub's'}^{t'} = x_{us'}^{t'} \cdot y_{ub'}^{t'}$ and denote $\mathcal{Z} = \{z_{ubs}\}$.
As a result of this migration and handover, the change of the E2E delay is
{\small{
$\widehat{\Delta d}_{ubb'ss'}^{t'}=\hat{d}_{ubs}^{t'} - \hat{d}_{ub's'}^{t'}$
}%
}%
, which takes into account both computation and communication aspects.
Throughout this paper, we use the symbol hat ($\hat{.}$) to denote estimation because $t'$ is a future time.
The total gain obtained from the migration and handover is thus defined by
%, but it involves a cost as the service downtime during migration.
%By considering the gain of reducing E2E delay, we can capture both computation and communication cost between running the service on the old and new servers, which is a common approach of the other works \cite{XiangSunICC2016}.
%We define the \textit{gain} $G$ of migration container from the server $s$ to $s'$, and/or handover from the BS $b_{b}$ to $b'$ as the total estimated E2E delay reduction before and after migration the service and/or handover to a new BS.
{\footnotesize{
\begin{align}
    G &= \sum_{b=1}^{|\mathcal{B}|}\sum_{s=1}^{|\mathcal{S}|}\sum_{u=1}^{|\mathcal{U}|}\sum_{b'=1}^{|\mathcal{B}|}\sum_{s'=1}^{|\mathcal{S}|}
    {\widehat{\Delta d}_{ubb'ss'}^{t'}  \hat{n}_{ub's'}^{t'}  z_{ubs}^{t}  z_{ub's'}^{t'}} \notag\\
    &= \sum_{u=1}^{|\mathcal{U}|}\sum_{b'=1}^{|\mathcal{B}|}\sum_{s'=1}^{|\mathcal{S}|} {\widehat{\Delta d}_{ubb'ss'}^{t'}  \hat{n}_{ub's'}^{t'}  z_{ub's'}^{t'}}\label{eq:gain2}
\end{align}
}%
}%
where $\hat{n}_{ub's'}^{t'}$ is the estimated total number of tasks that are offloaded from the MU $u$ via the new BS $b'$ to the new server $s'$ at the future time $t'$.
Since $z_{ubs}^{t}=1$ only for $b$ and $s$ that associate with the user $u$, we can reduce \eqref{eq:gain2} from the first equation to the second equation.
% which can capture a simple behavior of the MU.
%If we can predict accurately a moving path and behavior of the MU, we can estimate accurately the number of queries of the user at the future time $t'$.
%But this problem is out of the scope of the paper.
%We assume that the MUs only moves with a steady velocity, and without zigzag moving.
%Note that the typical MEC scenario is that a MU is running an application that continuously offloads tasks to a nearby server.
Note that if $u$ continuously offloads $n_{ubs}^{t}$ tasks to server $s$ via BS $b$, it will likely offload a similar number of tasks to the new server $s'$ via the new BS $b'$. Hence we assume $\hat{n}_{ub's'}^{t'} \approx {n}_{ubs}^{t}$.
%At the future time $t'$, if the total gain of reduction E2E delays of $\hat{n}_{u,b',s'}^{t'}$ queries is higher than the cost of the service downtime, the planner will make a migration decision to achieve the benefit of a closer edge server.
%After the service is migrated to the next server, the number of requests $n_{ubs}^{t}$ is reduced in half to keep both historical and updated demands of the MU.
%If the user moves faster and quickly changes to another BS, the number of queries is not increased high enough to compensate the cost of migration, resulting in discouragement of the migration.
%As a result, the number of requests $n_{ubs}^{t}$ discourages the migration.
%Subsequently, at the following BS, the estimated number of requests is likely small as well.
%So the system will not migrate to the next edge server because the gain is smaller than the cost of migration.

Besides the gain, migration and handover will also cause service interruption which can be measured by the total service downtime.
This downtime starts when $u$'s container $D_u$ becomes unavailable (due to connection being handed over from BS $b$ to $b'$, or container being migrated from server $s$ to $s'$), and ends when the service becomes available again.
Hence we denote it by $\widehat{DT}_{ubb'ss'}^{t'}$.
Thus, the total cost of the migration and handover is defined by
%Migrating the container reduces the E2E delay, but it involves a \textit{cost} $C$ due to service interruption, which is characterized by the total service downtime:
%We define the \textit{migration cost} for moving the container $D_u$ from the server $s$ to $s'$ at the future time $t'=t+\Delta t$ as follows:
{\footnotesize{
\begin{align}
\label{eq:cost}
    %C & =\sum_{b=1}^{|\mathcal{B}|}\sum_{s=1}^{|\mathcal{S}|} \sum_{u=1}^{|\mathcal{U}|} \sum_{b'=1}^{|\mathcal{B}|}\sum_{s'=1}^{|\mathcal{S}|}{\widehat{DT}_{ubb'ss'}^{t'}  z_{ubs}^{t}  z_{ub's'}^{t'}}\notag\\%
    C &=\sum_{u=1}^{|\mathcal{U}|} \sum_{b'=1}^{|\mathcal{B}|}\sum_{s'=1}^{|\mathcal{S}|}{\widehat{DT}_{ubb'ss'}^{t'}  z_{ub's'}^{t'}}.
\end{align}
}%
}%
%where $\widehat{DT}_{uss'bb'}^{t'}$ is the estimated total service downtime, which is the duration from the container $D_u$ becomes unavailable (due to connection handover from BS $b$ to $b'$, or migration from server $s$ to $s'$) until the service becomes available again.

%Thus, we formulate the problem of maximizing total profit which is the gain $G$ with the cost $C$ subtracted, as follows:
Our objective is to maximize total profit, defined as
{\footnotesize{
\begin{subequations}
	\begin{align}
	\label{eq:problemP1}
	%& \mathscr{P}1: \underset{z_{us'b'}^{t'}}{\text{argmax}}
	%\sum_{u=1}^{|\mathcal{U}|}\sum_{b'=1}^{|\mathcal{B}|}\sum_{s'=1}^{|\mathcal{S}|} ({\hat{E}_{ubb'ss'}^{t'}  \hat{n}_{ub's'}^{t'} - \widehat{DT}_{ubb'ss'}^{t'}  )z_{ub's'}^{t'}}
	& \arg\max_{z_{us'b'}^{t'}} (G-C),
	\end{align}
	\text{\small{ subject to:}}%
	\begin{align}
	%& \label{eq:constrainDomainZ} z_{ub's'}^{t'} \in \{0,1\}, ~ \forall s' \in \mathcal{S}, \forall b' \in \mathcal{B},\\
	& \label{eq:constrain1service}  \sum_{s'=1}^{|\mathcal{S}|}\sum_{b'=1}^{|\mathcal{B}|} z_{ub's'}^{t'} = 1,~\forall u \in \mathcal{U},\\
	& \label{eq:constrainResources} \sum_{u=1}^{|\mathcal{U}|} \sum_{b'=1}^{|\mathcal{B}|} \text{Res}(D_{u})  z_{ub's'}^{t'} \leq \text{Res}(s'), ~ \forall s' \in \mathcal{S}, \\
	& \label{eq:constrainRSSI} z_{ub's'}^{t'} = 0 ~ \text{if}~  \max{\{ \text{RSSI}_{ub'}^{t}, \widehat{\text{RSSI}}_{ub'}^{t'} \} } < \text{RSSI}_{min},\\
	& \label{eq:constrainMaxUserBTS} \sum_{u=1}^{|\mathcal{U}|} \sum_{s'=1}^{|\mathcal{S}|} z_{ub's'}^{t'} \leq N_{b'}, ~ \forall b' \in \mathcal{B}.
	\end{align}
\end{subequations}
}%
}%
The constraint (\ref{eq:constrain1service}) says that an offloaded service $D_u$ must be hosted by a single server, and a MU is associated with a single BS.
The constraint (\ref{eq:constrainResources}) means that the total resources required by $u$'s containers $D_u$, i.e., $\text{Res}(D_u)$, does not exceed the server's resources $\text{Res}(s')$, where the resources include CPU, memory, storage, and network I/O.
The constraint (\ref{eq:constrainRSSI}) imposes that the MU is not to be associated with a BS if either the measured RSSI or estimated RSSI
%\footnote{The measured RSSI samples are noisy and inconsistent \cite{HaratcherevChanelDynamic}, especially in a mobile environment. %, the channel conditions are changed rapidly due to interference, changing distance, fading.
%Hence, we used exponential moving average \cite{NadaPredictTriggerHandover2008} to reduce erratic, fluctuated signal.}
is below the minimum RSSI required by the receiver in order to decode signal.
The constraint (\ref{eq:constrainMaxUserBTS}) means that a BS $b'$ can serve a maximum number $N_{b'}$ of MUs.
%The last constraint (\ref{eq:constrainDomainZ}) defines the domain specific of the variable.

In the above, all the parameters at time $t$ are known and obtained by querying the central database.
For the parameters at time $t'$, how to estimate the change of E2E delay, $\widehat{\Delta d}_{ubb'ss'}^{t'}$ as in \eqref{eq:gain2}, and the service downtime, $\widehat{DT}_{ubb'ss'}^{t'}$ as in \eqref{eq:cost}, are presented in Section~\ref{subsec:gainE2EDelay} and Section~\ref{sec:orchestratedMechanism}, respectively.

The problem \eqref{eq:problemP1} is a \textit{multidimensional Knapsack problem} which is NP-hard \cite{frieze_approximation_1984}.
%Even there is no exact solution for the problem, the multi-constrained Knapsack problem is the well investigated problem.
%So we do no tend to propose an algorithm for the problem again.
Hence, we use the mixed-integer programming open-source solver CBC\cite{CBCSolver} to find a numerical solution.
In particular, we are interested in the case of $z_{ub's'}^{t'}=1$ which means that the user $u$ is connected to BS $b'$ and allocated an offloaded service to server $s'$ at the future time $t'$.
Based on our experimental observations, there always exists a feasible solution, i.e., migration-handover plan, for MUs.

%\subsection{End-to-End Delay Analysis}
\textbf{End-to-End delay analysis:}
\label{subsec:gainE2EDelay}
%--------------------------------------------------
%The E2E delay in this paper is defined as the period from the time when a MU sends a request to the edge computing system (more specifically an edge server) for offloading a task, until the time when the MU receives the task execution results from the system (the edge server).
At a future time $t'$, the estimated E2E delay $\hat{d}_{ubs}^{t'}$ consists of four components: processing, transmission, propagation, and queuing delays.
%\begin{equation}
%d_{u,b,s}^{t'} = d_{\text{proc},u,b,s}^{t'} + d_{\text{tran},u,b,s}^{t'} + d_{\text{prop},u,b,s}^{t'} + d_{\text{queue},u,b,s}^{t'}
%\end{equation}
The \textit{processing delay} is the duration of executing an offloading task on the edge server which depends on the computational power of the server, and can be estimated as $\hat{d}_{\text{proc},ub's'}^{t'} \approx d_{\text{proc},ubs}^{t} \cdot {C_s}/{C_{s'}}$,
where $C_s, C_{s'}$ are the computational power of the source server $s$ and that of the destination server $s'$, respectively, and can be measured using benchmarking software\footnote{One such example is {\tt sysbench}, and it needs to run once on each server.}.
The \textit{transmission delay} is the time to transmit the offloading task and receive the task execution result, and can be estimated as $\hat{d}_{\text{tran},ubs}^{t'} \approx {\hat{S}_{u}^{t'}}/{\hat{B}_{ubs}^{t'}},$
where $\hat{S}_{u}^{t'}={S}_{u}^{t}$ is the (estimated) task size, $\hat{B}_{ubs}^{t'}=\min{(\hat{B}_{ub}^{t'}, \hat{B}_{bs}^{t'})}$ is the estimated bandwidth.
The bandwidth between the MU and BS $\hat{B}_{ub}^{t'}$ can be estimated based on the $\widehat{\text{RSSI}}_{ub}^{t'}$ and the access wireless technology \cite{MapRSSIRateIEEE80211n}, and the bandwidth between BS and server $\hat B_{bs}^{t'}=B_{bs}^{t}$ can be obtained by querying the central database. In estimating $\hat{d}_{\text{tran},ubs}^{t'}$, we ignore the transmission delay of the task execution result due to its much smaller size compared to the task itself.
The \textit{propagation delay} is the round-trip time (RTT) of propagating a single bit between the MU and the server. It includes two parts: (1) between the MU and the BS via wireless, which can be neglected due to the very short distance as compared to the speed of light, and (2) between the BS and the server via wired network, which can be far away from each other, as is denoted by $\text{RTT}_{bs}$.
For \textit{queuing delay}, we assume for simplicity that it remains the same before and after migration.
Finally, the change of E2E delay is estimated as:
{\footnotesize{
\begin{equation}
\begin{aligned}
\label{eq:deltaE2EDelayNew}
\widehat{\Delta d}_{ubb'ss'}^{t'} &= \hat{d}_{ubs}^{t'} - \hat{d}_{ub's'}^{t'}
 \approx d_{\text{proc},ubs}^{t} \Big( 1 - \frac{C_s}{C_{s'}} \Big) + \\
& S_{u}^{t} \Big(  \frac{1}{\hat{B}_{ubs}^{t'}} -  \frac{1}{\hat{B}_{ub's'}^{t'}} \Big) +
 \big({\text{RTT}}_{bs} - {\text{RTT}}_{b's'} \big)
\end{aligned}
\end{equation}
}}%
We note that it could be negative if the migration decision is not made properly.

%----------------------------------------------
%\subsection{Mechanism Part B: Timing}
\subsection{Best Triggering Time}
\label{sec:orchestratedMechanism}
%----------------------------------------------
This subsection determines the best time to trigger BS handover and the best time to trigger container migration.
We propose a technique called \textit{delta checkpoint} to perform container migration for stateful applications, which consists of two phases:
\begin{itemize}
    \itemsep0em
    \item Pre-migration phase: in this phase, we checkpoint (i.e., snapshot the memory of) the current container and transfer the whole memory state to the destination server, while leaving the container continue to run.
    \item Migration phase: we checkpoint the container again and save the difference between this and the previous checkpoint as a {\em delta memory state}, which is much smaller than the memory state in the pre-migration phase and is transferred to the destination server.
\end{itemize}
Note that in the delta checkpoint technique, we assume that base container images\footnote{A container image is an immutable file that contains a snapshot of a container.} are available at the source and destination servers (which can be done by downloading in advance).
Recall that at a future time $t' = t + \Delta t$, a MU $u$ moves from the current BS $b$ to the vicinity of another BS $b'$.
%the best offloading edge server is assumed the server $s'$.
We need to estimate the time taken for migrating $u$'s container from server $s$ (which is connected to $b$) to $s'$ (which is connected to $b'$). %(Section~\ref{subsec:estimateMigrationTime}).
This total migration time is a sum of pre-migration time and migration time, i.e.,
{\small{
\[ \hat{T}_{\text{total-mig},uss'}^{t'} = \hat{T}_{\text{pre-mig},uss'}^{t'}+\hat{T}_{\text{mig},uss'}^{t'},\]
}%
} where:
{\footnotesize{
\begin{align}
\label{eq:preMigrationTime}
\hat{T}_{\text{pre-mig},uss'}^{t'} &= \hat{T}_{\text{chkpt},us}^{t'} + \hat{T}_{\text{pre-trans},uss'}^{t'}, \\
\label{eq:migrationNonLiveMigration}
\hat{T}_{\text{mig},uss'}^{t'} &= \hat{T}_{\text{chkpt},us}^{t'} + \hat{T}_{\text{trans},uss'}^{t'} + \hat{T}_{\text{restore},us'}^{t'}.
\end{align}
}%
}%
\iffalse
In the above, $\hat{T}_{\text{chkpt},us}^{t'}, \hat{T}_{\text{restore},us'}^{t'}$ are the checkpoint time and restore time which depend on the size of the container image $D_u$ and computational capacity of the servers;
$\hat{T}_{\text{pre-trans},uss'}^{t'}, \hat{T}_{\text{trans},uss'}^{t'}$ are the time to transfer pre-migration's checkpointed files and the time to transfer the delta checkpointed files, which depends on the size of files and the network bandwidth between the two servers.
\else
In the above,
\begin{itemize}
    \item $\hat{T}_{\text{chkpt},us}^{t'} = \psi_{s}^{t'}  {S_{D_u}}/{C_{s}}$ is the checkpoint time which depends on the size of the container image $D_u$ and computational power of the source server;

    \item $\hat{T}_{\text{pre-trans},uss'}^{t'} = {\hat{S}_{\lambda_u}^{t'}}/{B_{ss'}^{t'}}$ is the time to transfer pre-migration's checkpointed files with the estimated size $\hat{S}_{\lambda_u}^{t'}$ over the network bandwidth $B_{ss'}^{t'}$ between the two servers;

    \item $\hat{T}_{\text{trans},uss'}^{t'} = {\hat{S}_{\Delta \lambda_u}^{t'}}/{B_{ss'}^{t'}}$ is the time to transfer $\hat{S}_{\Delta \lambda_u}^{t'}$ which is the estimated size of delta memory state between the migration's checkpointed files and the pre-migration's checkpointed files;

    \item $\hat{T}_{\text{restore},us'}^{t'} = \rho_{s'}^{t'}  ({S_{D_u} + \hat{S}_{\lambda_u}^{t'} + \hat{S}_{\Delta \lambda_u}^{t'}})/{C_{s'}}$ is the time to restore the migrated container at the destination server.
\end{itemize}
The parameters $\psi_{s}^{t'}$ and $\rho_{s}^{t'}$ can be inferred by using the historical information of checkpoints and restores of all containers hosted on the server $s$.

Estimating the size of delta memory state $\hat{S}_{\Delta \lambda_u}^{t'}$ is hard because it varies substantially between different computation tasks, making static information (e.g., container image size) much less instrumental.
To solve this problem, we use a heuristic technique as follows. After the container $D_u$ processes the MU $u$'s tasks for a certain period of time, say at the time $t_0$, the hosting edge server triggers two consecutive checkpoints to $D_u$ while leave it running. Then the server measures the size of the two checkpointed files, where the first is $S_{\lambda_u}^{t_0}$ and the difference between the two is the delta memory state $S_{\Delta \lambda_u}^{t_0}$.
Thus, we estimate the size of pre-migration checkpoint and the size of delta memory state to be $\hat{S}_{\lambda_u}^{t'} \approx S_{\lambda_u}^{t_0}$ and $\hat{S}_{\Delta \lambda_u}^{t'} \approx S_{\Delta \lambda_u}^{t_0}$, respectively.
\fi

The time to handover the connection from the BS $b$ to $b'$, which we denote by $t_{\text{ho},ubb'}$, can be estimated using the relative RSSI hysteresis technique \cite{NadaPredictTriggerHandover2008}.

\begin{figure}[ht]
    \centering
    \includegraphics[width=0.95\linewidth]{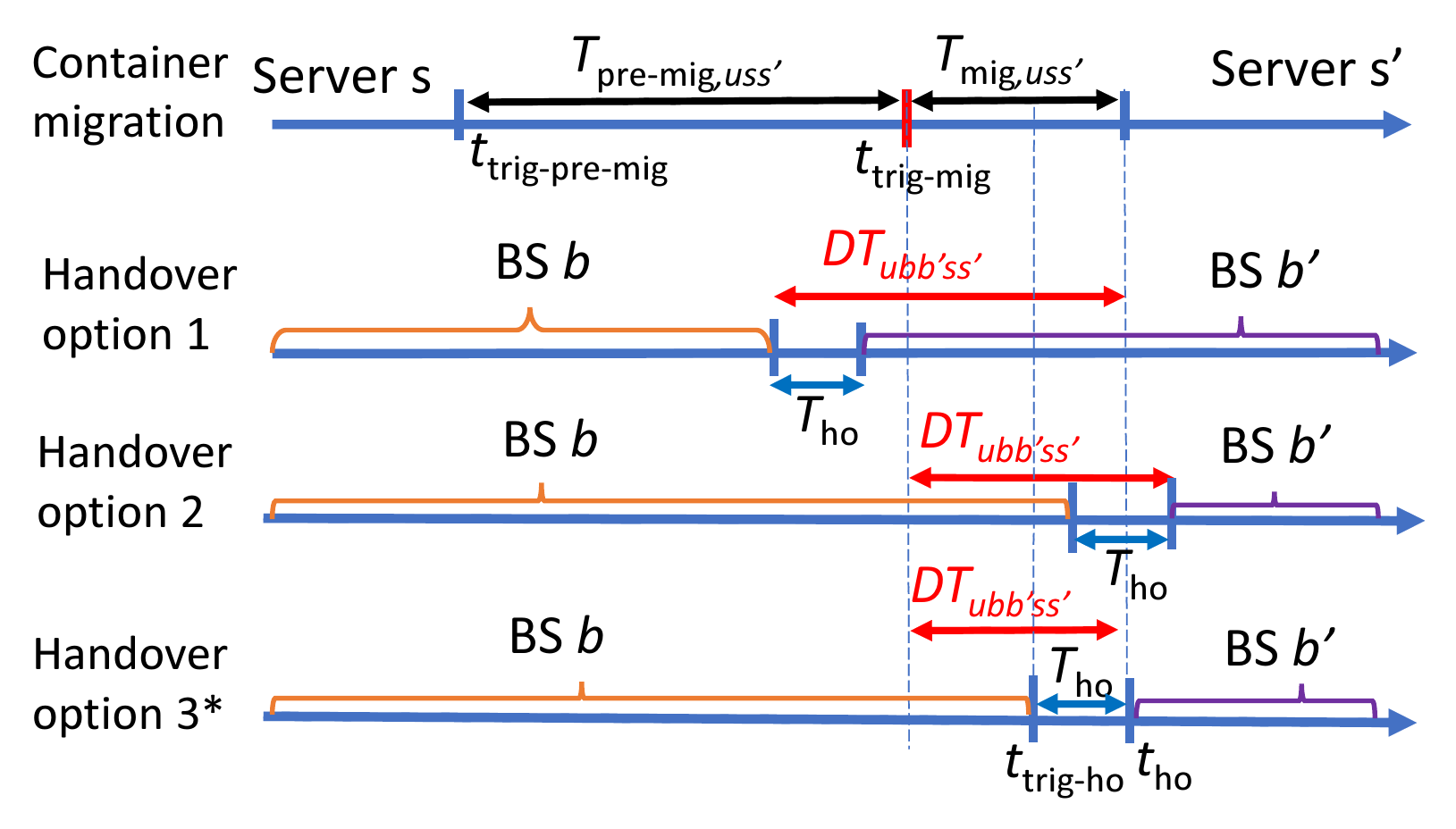}
    \caption{Handover options w.r.t the migration timeline.}
    	%The timeline of container migration versus the timeline of handover.}
    %		Different handover timelines corresponding to the container migration timeline (top).}
    \label{fig:timeSwitchingAP}
\end{figure}
Based on the above estimates, we explain how to trigger handover and container migration in an orchestrated manner in order to minimize total service downtime.
When the MU $u$ moves from the BS $b$ to $b'$, the container $D_u$ is either migrated to another server $s'$, or still running at the old server $s$ that can be considered as a special case of the former when $s'=s$.
%We design a general triggering mechanism for both cases by considering the latter case as a special case of the former when $s'= s$.
The order of triggering handover and migration processes can significantly affect the total service downtime. As shown in Fig. \ref{fig:timeSwitchingAP},
%If the two processes happens consecutively or separately, the total service downtime is at least equal to the sum of migration time and handover time.
the first timeline describes the container migration process including pre-migration and migration.
The second and third timelines describe early and late triggering of handover, respectively, and it shows that the total service downtime is larger than the maximum of the migration time and the handover time ($T_{\text{ho},ubb'}$).
Only the fourth timeline presents the best time to trigger handover, which achieves the minimal total service downtime as is estimated to be
{\footnotesize{
\[ \widehat{DT}_{ubb'ss'}^{t'} = \max{ \{ \hat{T}_{\text{mig},uss'}^{t'}, T_{\text{ho},ubb'} \} }. \]
}%
}%
%As shown in Fig. \ref{fig:timeSwitchingAP}, in order to achieve the orchestrated migration-handover mechanism, the time
So, in our proposed migration-handover mechanism, the time to trigger pre-migration, migration and handover, with a 10\% error margin, is given as ${t}_{\text{trig-pre-mig}}, {t}_{\text{trig-mig}}$ in the first timeline, and ${t}_{\text{trig-ho}}$ in the fourth timeline as shown in Fig.~\ref{fig:timeSwitchingAP}.

%----------------------------------------------
\section{Performance Evaluation}
\label{sec:experimentalSetup}
%----------------------------------------------

%We have developed a MEC testbed. It has a cloud layer which consists of a central controller and a cloud server, and an edge layer which consists of three edge servers associated with three WiFi access points serving as three BSs.

We implement container deployment using Docker \cite{docker} which is widely adopted in the industry. We implement container migration based on our proposed delta checkpoint technique, using an open source tool CRIU \cite{criu}.
% to mimic the principle of container live migration \cite{Mirkin08containerscheckpointing}.

%----------------------------
\subsection{Experiment Setup}
\begin{figure}[!t]
	\centering
	\includegraphics[width=0.92\linewidth]{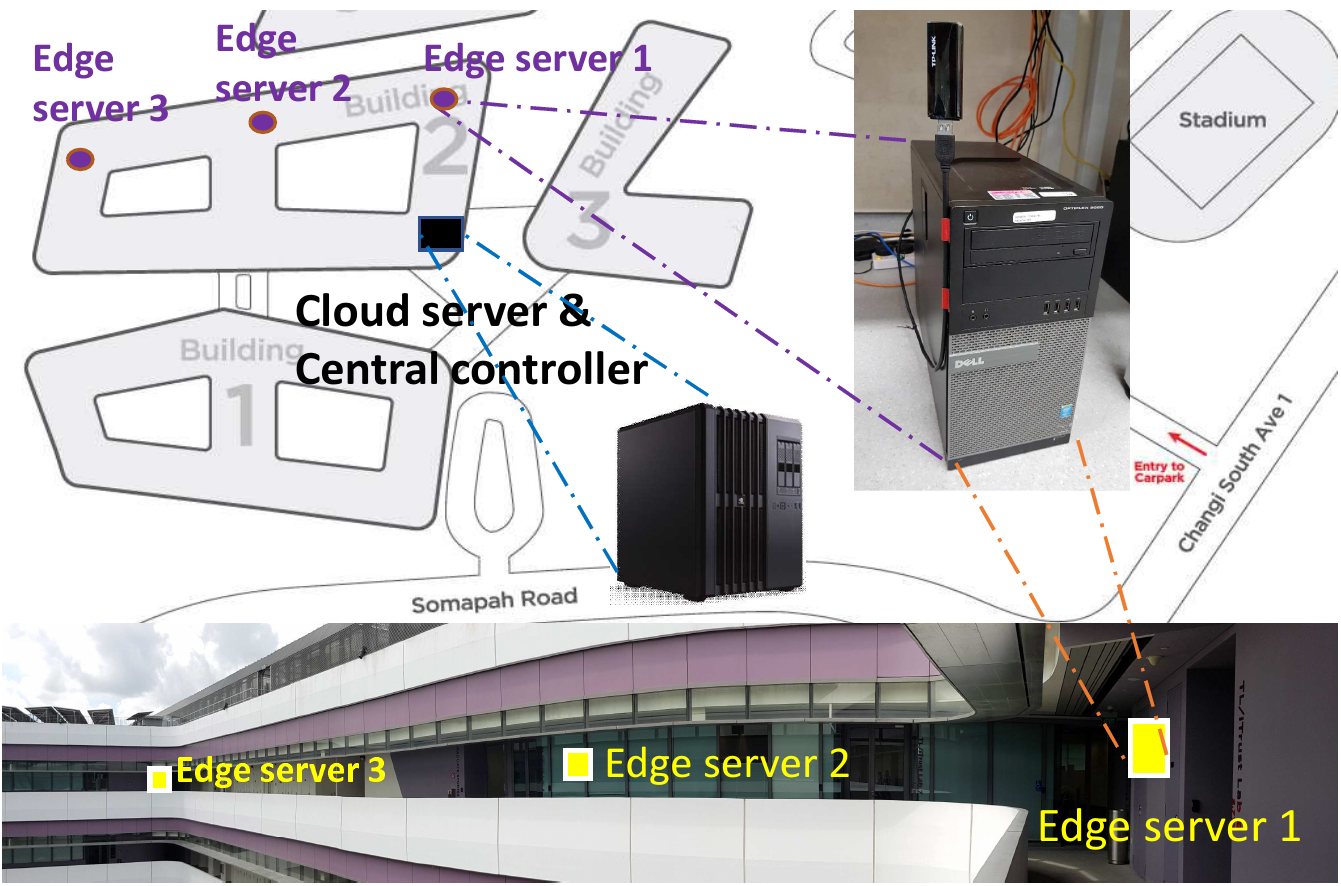}%scenario.eps}
	%\caption{Our testbed consists of three edge servers and one cloud server.}
    \caption{Layout of Testbed.}
	\label{fig:scenario}
\end{figure}

%----------------------------
Fig.~\ref{fig:scenario} shows our testbed of the presented MEC system, which includes a cloud server which also runs a central controller, and three edge servers.
%The three edge servers are deployed on a linear topology at the coordinates $30$m, $100$m, and $170$m.
%The servers are installed with Ubuntu 16.04 (kernel 4.13.0), CRIU \cite{criu} version 3.8.1, Docker \cite{docker} version 17.03.2-ce.
%Each of the three edge servers runs on a 4-core Intel\textsuperscript{\textregistered} Core\texttrademark~i7-4790 3.60 GHz and 8 threads with 16 GB RAM, and the cloud server runs on a 6-core Intel\textsuperscript{\textregistered} Core\texttrademark~i7-6850K 3.60 GHz and 12 threads with 128 GB RAM.
Each of the three edge servers runs on a 4-core Intel Core i7-4790 3.60 GHz (8 threads) with 16 GB RAM, and the cloud server runs on a 6-core Intel Core i7-6850K 3.60 GHz (12 threads) with 128 GB RAM.
We use Linux traffic control tool, \texttt{tc}, to emulate WAN connections~\cite{akamaiQ1RP2017}, where the connection between any two adjacent edge servers is configured as 100 Mbps bandwidth and 50ms latency~\cite{Lele17}, and the connection between each edge server and the cloud server is configured as 75 Mbps bandwidth and 150ms latency~\cite{Chen_SEC2017}.
%\iffalse
%\fi

We implement three applications (available at \textit{https://gitlab.com/ngovanmao/edgeapps}) to be the services for offloading image processing tasks, and package each into a Docker container:
%\iffalse %TODO: out of space
\begin{itemize}
	\itemsep0em
	\item a face recognition service based on \textit{Openface} \cite{amos2016openface},
	\item an object recognition service based on \textit{Yolo} \cite{YOLOv3OpenCV}, and
	\item a \textit{Simple} service which is a dumb TCP server that simply responds to each incoming offloading request with an incrementing counter (and hence the processing delay is treated as zero).
\end{itemize}
%\else
\iffalse
(i) a face recognition service based on \textit{Openface} \cite{amos2016openface},
(ii) an object recognition service based on \textit{Yolo} \cite{YOLOv3OpenCV}, and
(iii) a \textit{Simple} service which is a dumb TCP server that simply responds to each incoming offloading request with an incrementing counter (and hence the processing delay is treated as zero).
\fi

To simulate stateful applications, all the three applications store and increment counter after each incoming offloading request.
The counter is checked before and after migration to ensure consistent state of each offloaded service.
%Each service runs in a different container, and henceforth we refer to it as a service or a container interchangeably.
%The specifications of the evaluated services are described in Table \ref{tab:SpecificationService}.

In order to make our experimental results reproducible, we develop a \textit{virtual MU} in a simulated mobility environment rather than testing an actual smartphone in an actual mobility environment which is subject to many uncontrollable factors. %, such as a smartphone which can produce inconsistent results due to different app running status and software/hardware configurations.
Our virtual MU has all the required functional features and moving behavior of an actual MU. It offloads computational tasks (i.e., image processing) to one of the three real edge servers or a cloud server that we deploy as in Fig.~\ref{fig:scenario}. %, i.e, discovery, handover service, resource monitor, and application services.
%Besides these features, we simulate the MUs with moving direction and velocity.
To simulate the handover behavior, we run the virtual MU on a separate computer and use \texttt{iptable} to specify the single-hop traffic path between MU and its associated BS.
We also configure each WiFi AP as a network address translation (NAT) router to specify the single-hop traffic path between the MU's associated BS and the MU's offloading server. %to ensure the same forward and backward paths of data traffic from the MU to an edge server via its associated BS.
%The BSs are collocated with the edge servers

The MU uses the path loss model \cite{Rappaport01book} to generate RSSI values which will be used to trigger handover.
The handover time of the virtual MU is set to 500ms \cite{Mishra_Empirical80211Handoff_Sigcom2003}.
To simulate moving, %we let the MU continuously make round-trips between the starting point of coordinate 0 and the end point of coordinate 200\,m.
we let MU continuously make round-trips between the starting point and the end point, as shown in Fig. \ref{fig:scenario}.
The velocity is set to 0.5\,m/s in the cases of Openface and Yolo, and 1\,m/s in the case of Simple.  We run each experiment for 1600s.
%The simulated MU using the Simple service moves with velocity 1m/s, while the simulated MU using Openface or Yolo service moves with velocity 0.5 m/s since the migration time is much higher than the Simple service.

We implement four \textit{Planner} modules under the \textit{Central Controller} (see Fig.~\ref{fig:SystemArch}) for comparison:
\iffalse
\begin{itemize}
\itemsep0em
\item \textit{Cloud planner:} always allocates a container to the cloud (i.e., no migration) regardless of location of MUs.
\item \textit{Random planner:} allocates a container to a randomly-chosen server.
%During handover, the disconnection event of the MU triggers container migration to a random server.
% which can be the same or a different server.
%, and handles handover based on RSSI-threshold technique.
\item \textit{Nearest planner:} allocates a container to the nearest server of the MU.
%, and handles handover based on RSSI-threshold technique.
%During handover, the disconnection event of the MU triggers container migration to the next nearest server.
\item \textit{Orchestrated planner:} allocates a container and handover BS connection using our proposed mechanism (Section \ref{sec:DesignCoordinatedMechanism}).
% orchestrated container migration and handover connection with optimization placement
%This planner makes a handover decision based on RSSI-hysteresis and collaboratively triggers pre-migration and migration.
\end{itemize}
\else
%\iffalse
(1) \textit{Cloud planner:} always allocates a container to the cloud (i.e., no migration) regardless of location of MUs.
(2) \textit{Random planner:} allocates a container to a randomly-chosen server.
(3) \textit{Nearest planner:} allocates a container to the nearest server of the MU.
(4) \textit{Orchestrated planner:} allocates a container and handovers BS connection using our proposed mechanism (Section \ref{sec:DesignCoordinatedMechanism}).
\fi
The first three planners are triggered when BS handover is triggered due to low RSSI signal.
%----------------------------------------------
\subsection{Experiment Results}
\label{sec:expResults}
%----------------------------------------------

We evaluate the above MEC system in terms of two performance metrics: E2E delay and total service downtime experienced by MU.

%----------------------------------------------
\subsubsection{End-to-end delay}
\label{subsec:ResultE2Edelay}

\begin{figure*}
	\centering
	\begin{minipage}{0.75\linewidth}
		\centering
		\subfloat[Simple service]{
			\includegraphics[width=0.33\linewidth]{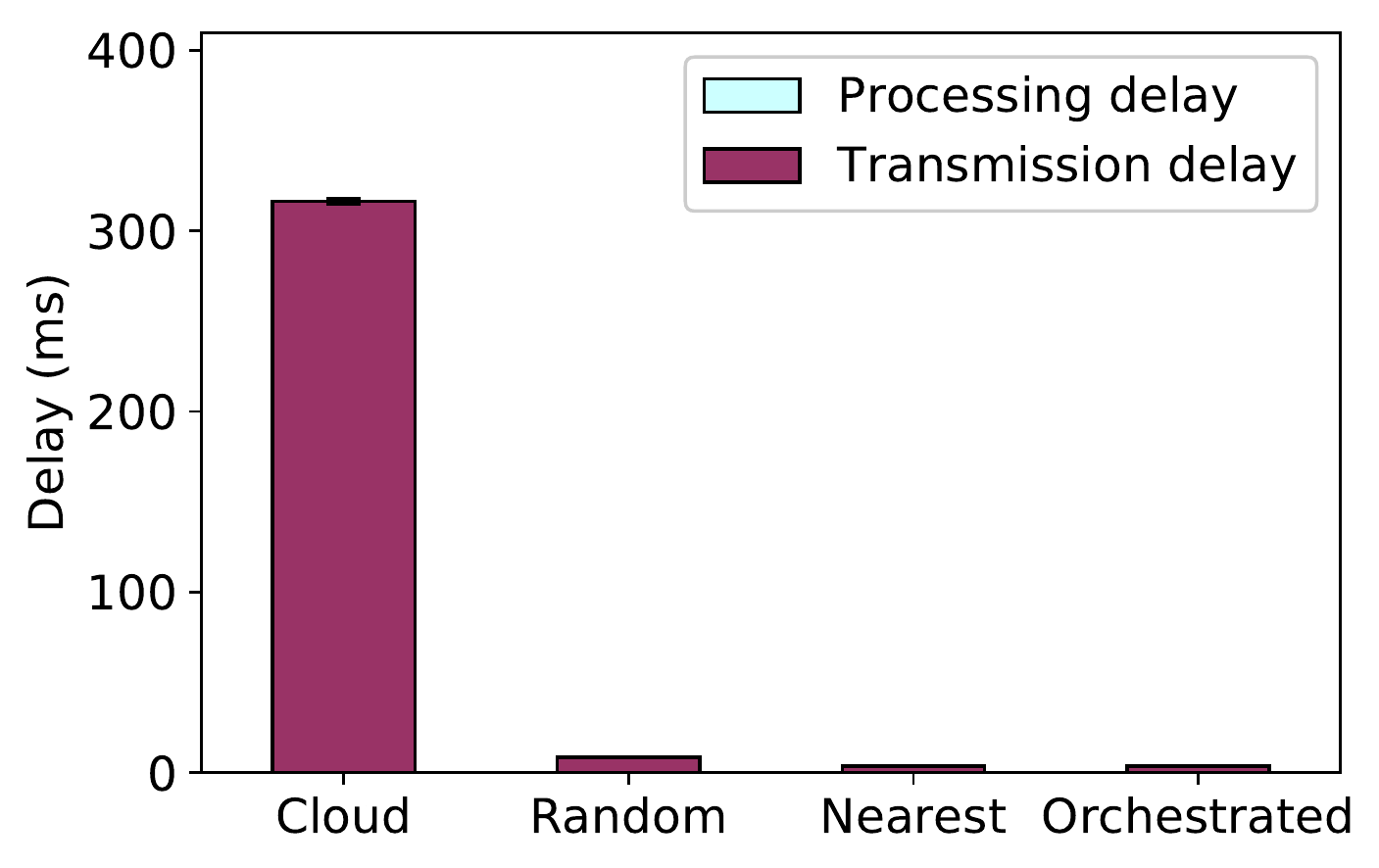}%
			\label{fig:E2EDelaySimple}%
		}\!\!\!%
		\subfloat[Openface service]{
			\includegraphics[width=0.33\linewidth]{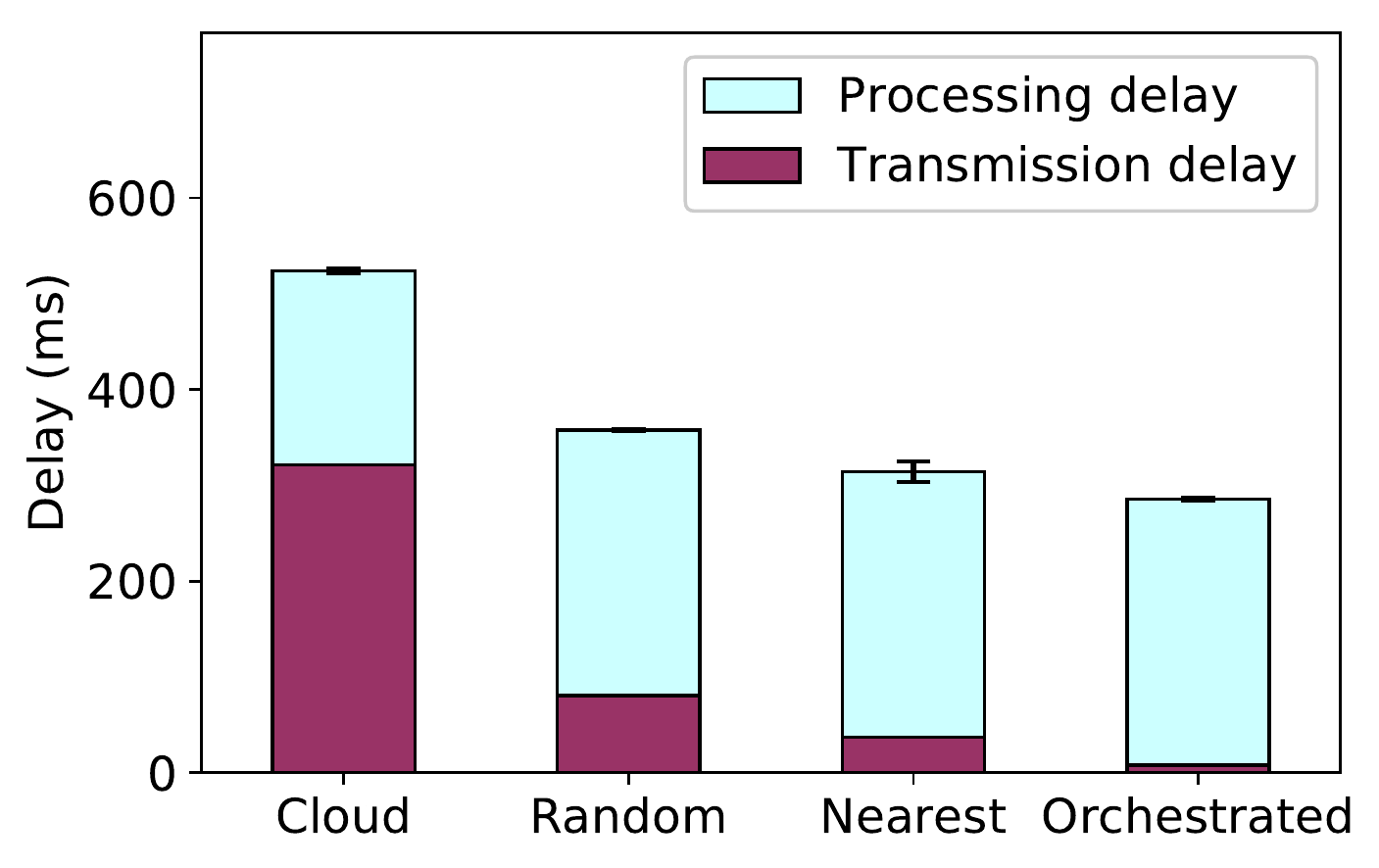}%
			\label{fig:E2EDelayOpenface}%
		}\!\!\!%
		\subfloat[Yolo service]{
			\includegraphics[width=0.33\linewidth]{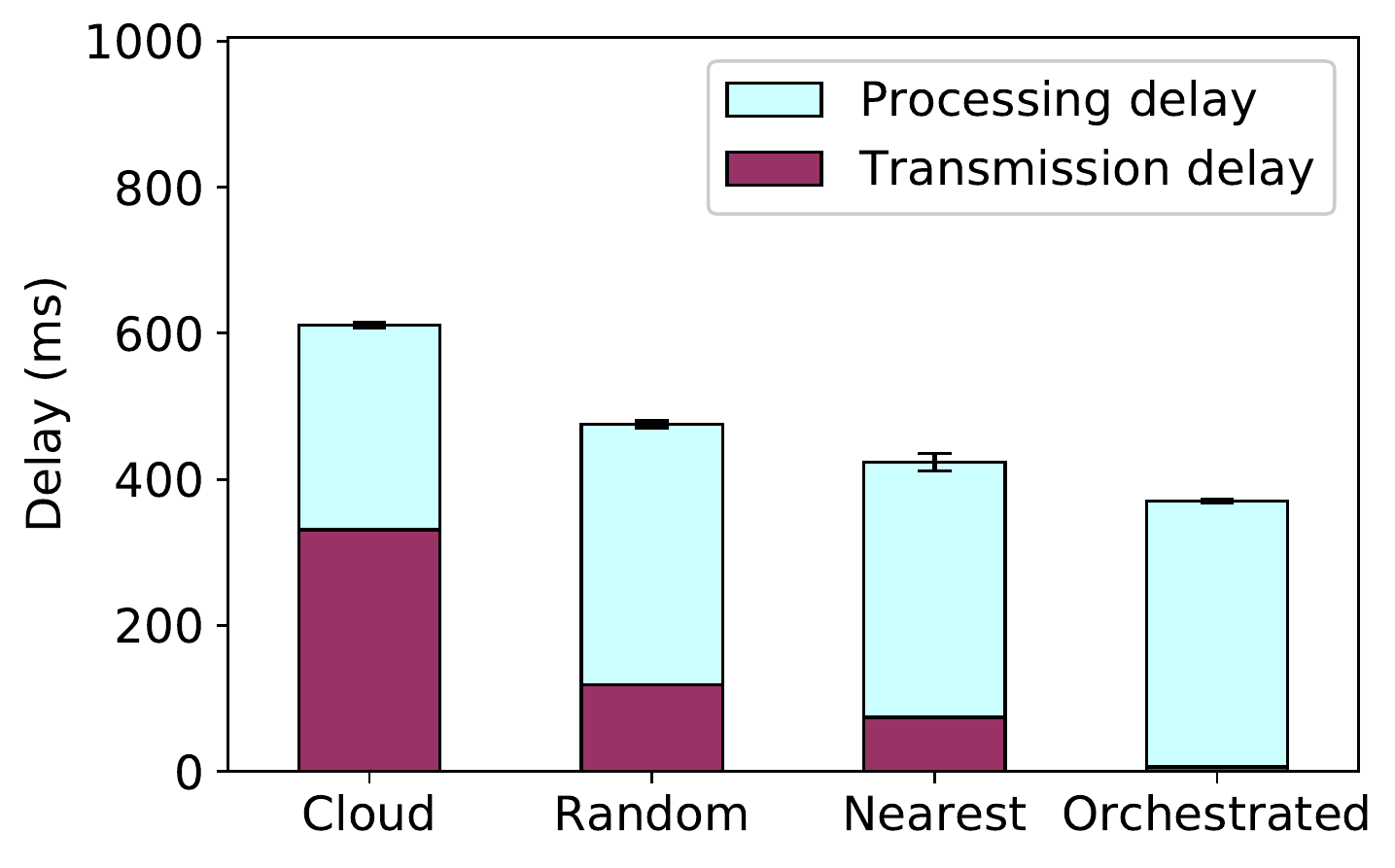}%
			\label{fig:E2EDelayYolo}%
		}\!\!\!%
		\caption{End-to-end (E2E) delay breakdown (which consists of mean processing delay and mean transmission delay) of a MU who offloads tasks to one of the three offloaded services under evaluated planners. The error bars indicate the E2E delay standard deviation.}
		\label{fig:E2EDelays}
	\end{minipage}~~~%
	\begin{minipage}{.25\linewidth}
		\centering
		\includegraphics[width=1.0\linewidth]{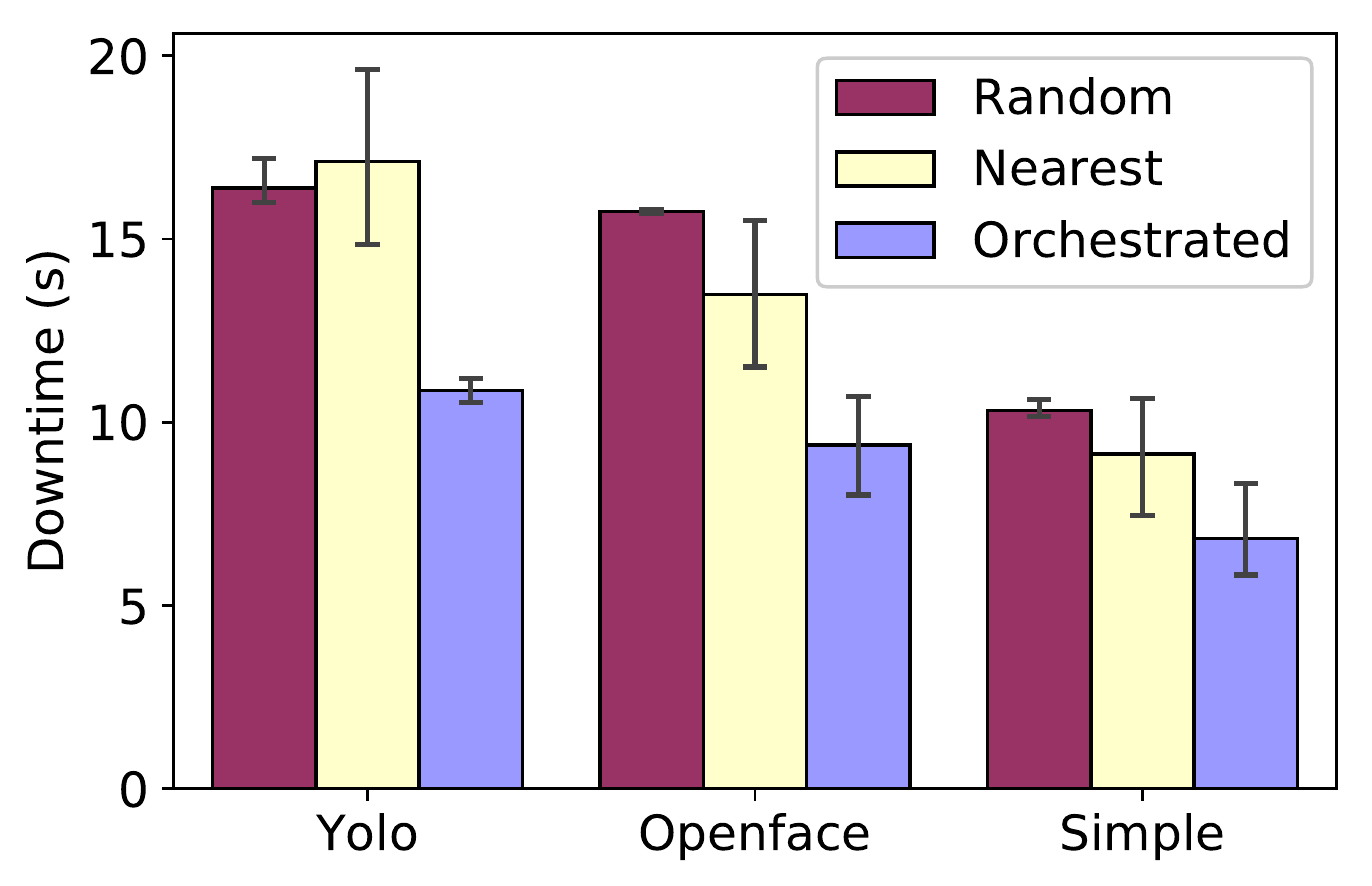}
		\caption{Total service downtime of a MU who offloads tasks to one of three offloaded services under three migration planners.}
		\label{fig:DowntimeCompared4planners}
	\end{minipage}
\end{figure*}

First, we show the statistical results of E2E delay (which consists of mean processing delay and mean transmission delay) of a MU who offloads tasks to one of the three offloaded services under four evaluated planners in Fig.~\ref{fig:E2EDelays}.
As we can see, the cloud planner incurs significant high E2E delay in which the long transmission delay dominates the short processing delay.
In Fig.~\ref{fig:E2EDelaySimple}, the E2E delay of \textit{Simple} service reflects the network configurations of the testbed since there is no processing delay.
%despite the cloud planner obtains the smallest processing time due to higher computation power of the cloud server, it incurs much longer transmission delay which results in the highest E2E delay.
As shown in Figs.~\ref{fig:E2EDelayOpenface}, \ref{fig:E2EDelayYolo}, the processing delay under the random, nearest, and orchestrated planners are more or less the same, but the transmission delay under the orchestrated planner is significantly lower than that under the random and nearest planners.
For example, for \textit{Openface} service, the transmission delay under the orchestrated planner is just 10\%, 22\% of that under the random and nearest planners, respectively.
%But the transmission delays under the random planner for \textit{Openface} and \textit{Yolo} services are 54\% and 38\% larger than those under the nearest planner, respectively.
%The transmission delay under the orchestrated planner is likely negligible, specifically those value for \textit{Simple}, \textit{Openface}, and \textit{Yolo} services are $3.7$\,ms, $8.3$\,ms, $6.1$\,ms, respectively.
As a result, the MU under the orchestrated planner achieves the lowest E2E delay.
For \textit{Yolo} service, the E2E delay under the orchestrated planner is 22.2\% and 12.6\% lower than that of the random and nearest planners, respectively.

\iffalse
\begin{figure*}[!htb]
	\centering
	\subfloat[Simple service]{
		\includegraphics[width=0.33\linewidth]{\images/delay_simple_stats.eps}%
		\label{fig:E2EDelaySimple12}%
	}\!\!\!%
	\subfloat[Openface service]{
		\includegraphics[width=0.33\linewidth]{\images/delay_openface_stats.eps}%
		\label{fig:E2EDelayOpenface12}%
	}\!\!\!%
	\subfloat[Yolo service]{
		\includegraphics[width=0.33\linewidth]{\images/delay_yolo_stats.eps}%
		\label{fig:E2EDelayYolo12}%
	}\!\!\!%
	\caption{End-to-end (E2E) delay breakdown (which consists of mean processing delay and mean transmission delay) of a MU who offloads tasks to one of the three offloaded services under four evaluated planners. The error bars indicate 95\% confidence interval of the E2E delay.}
	\label{fig:E2EDelays12}
\end{figure*}
\fi

\begin{figure*}[!htb]
	\centering
	\subfloat[Cloud planner]{
		\includegraphics[width=0.25\linewidth]{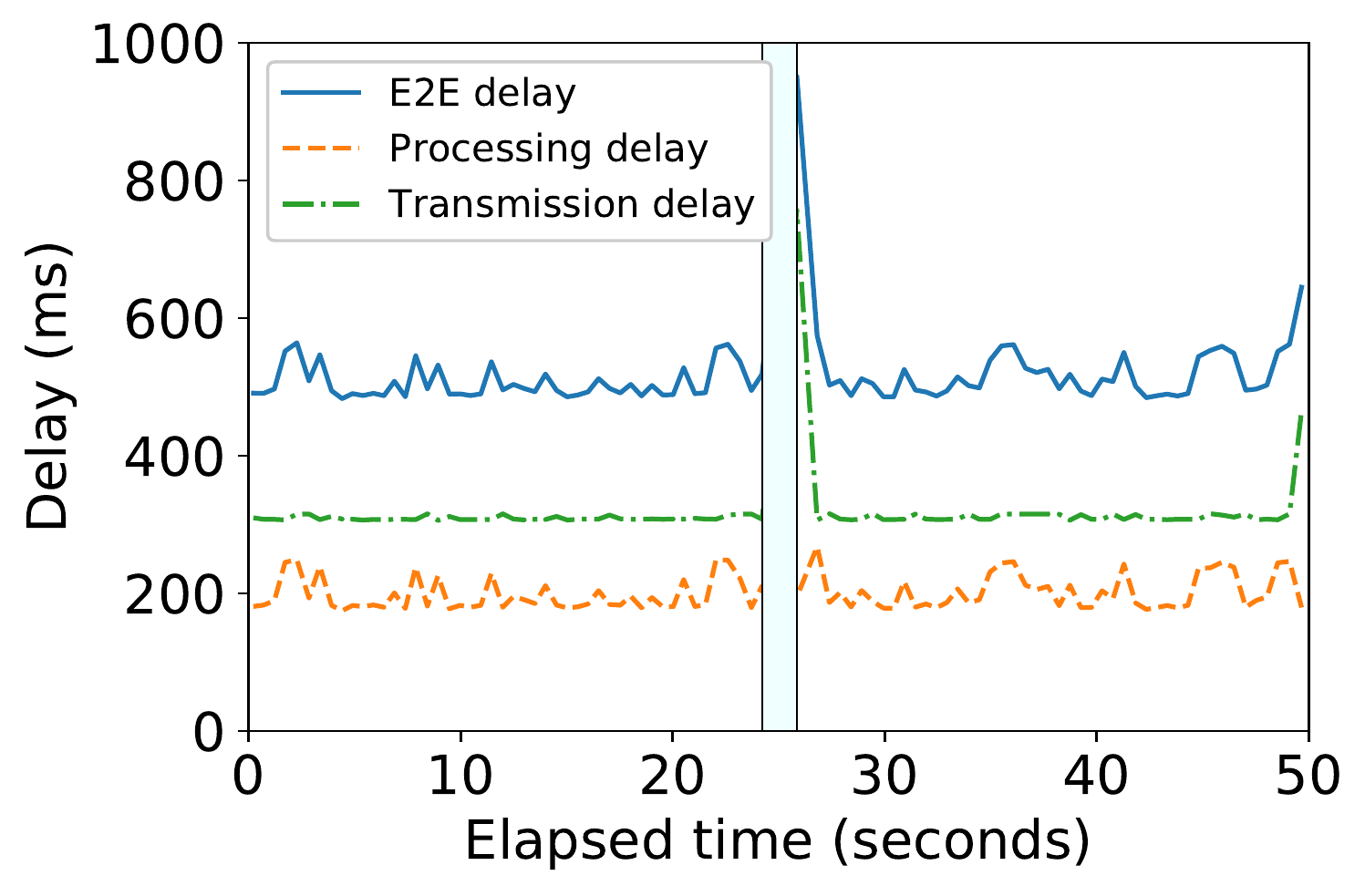}
		\label{fig:E2EdelayOpenfaceSimCloud}%
	}\!\!\!%
	\subfloat[Random planner]{
		\includegraphics[width=0.25\linewidth]{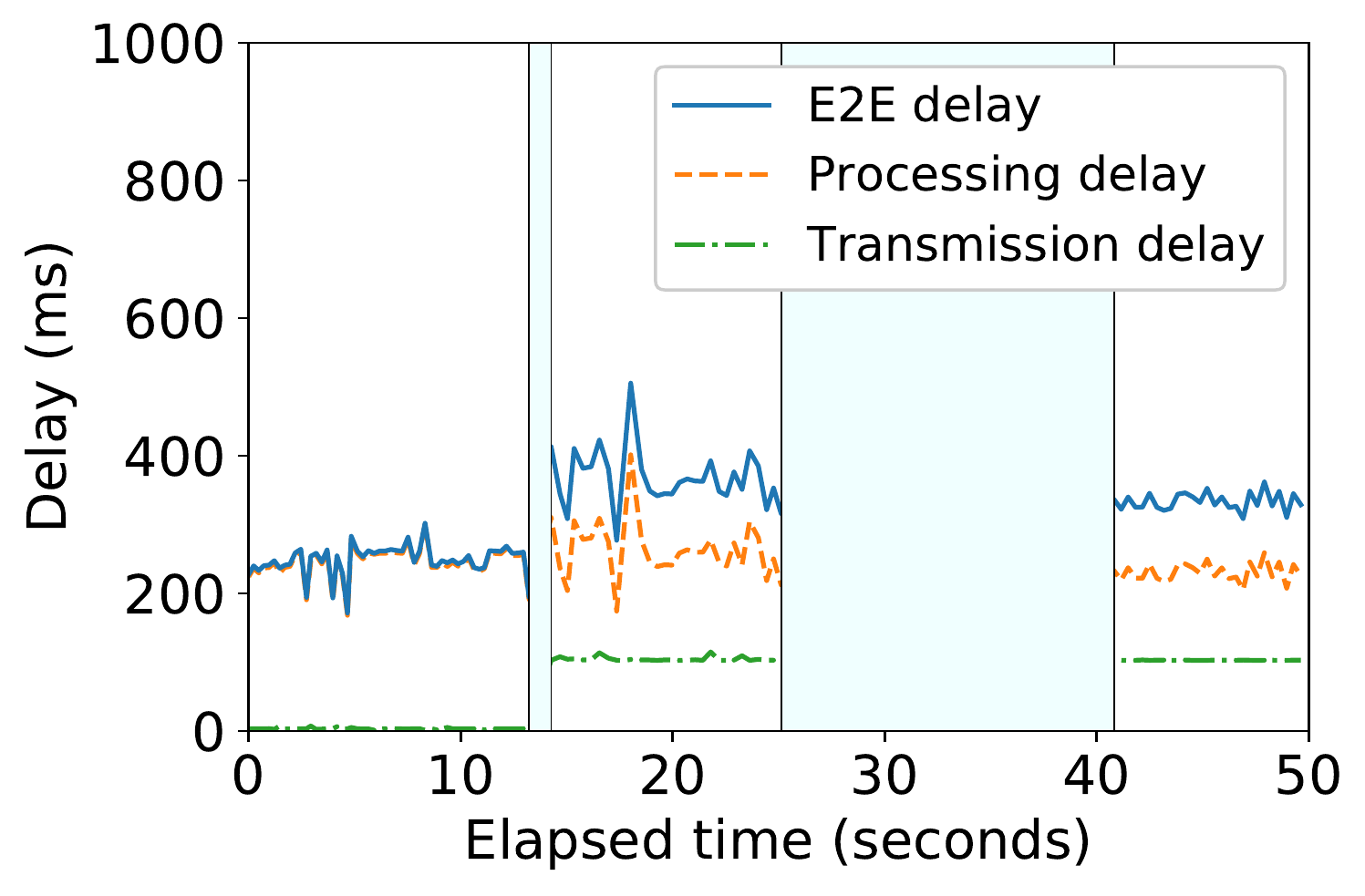}
		\label{fig:E2EdelayOpenfaceSimRandom}%
	}\!\!\!%
	\subfloat[Nearest planner]{
		\includegraphics[width=0.25\linewidth]{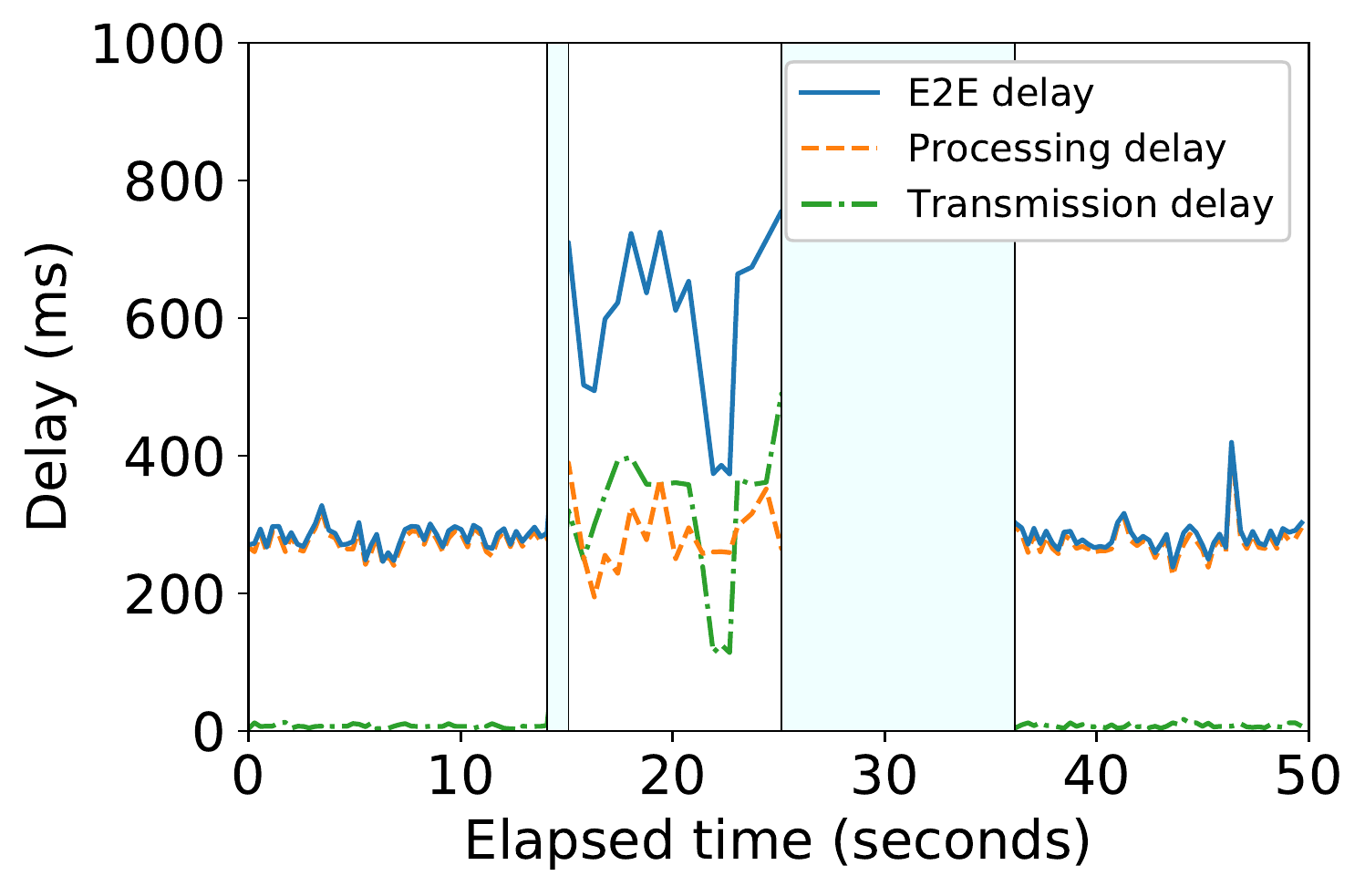}
		\label{fig:E2EdelayOpenfaceSimNearest}%
	}\!\!\!%
	\subfloat[Orchestrated planner]{
		\includegraphics[width=0.25\linewidth]{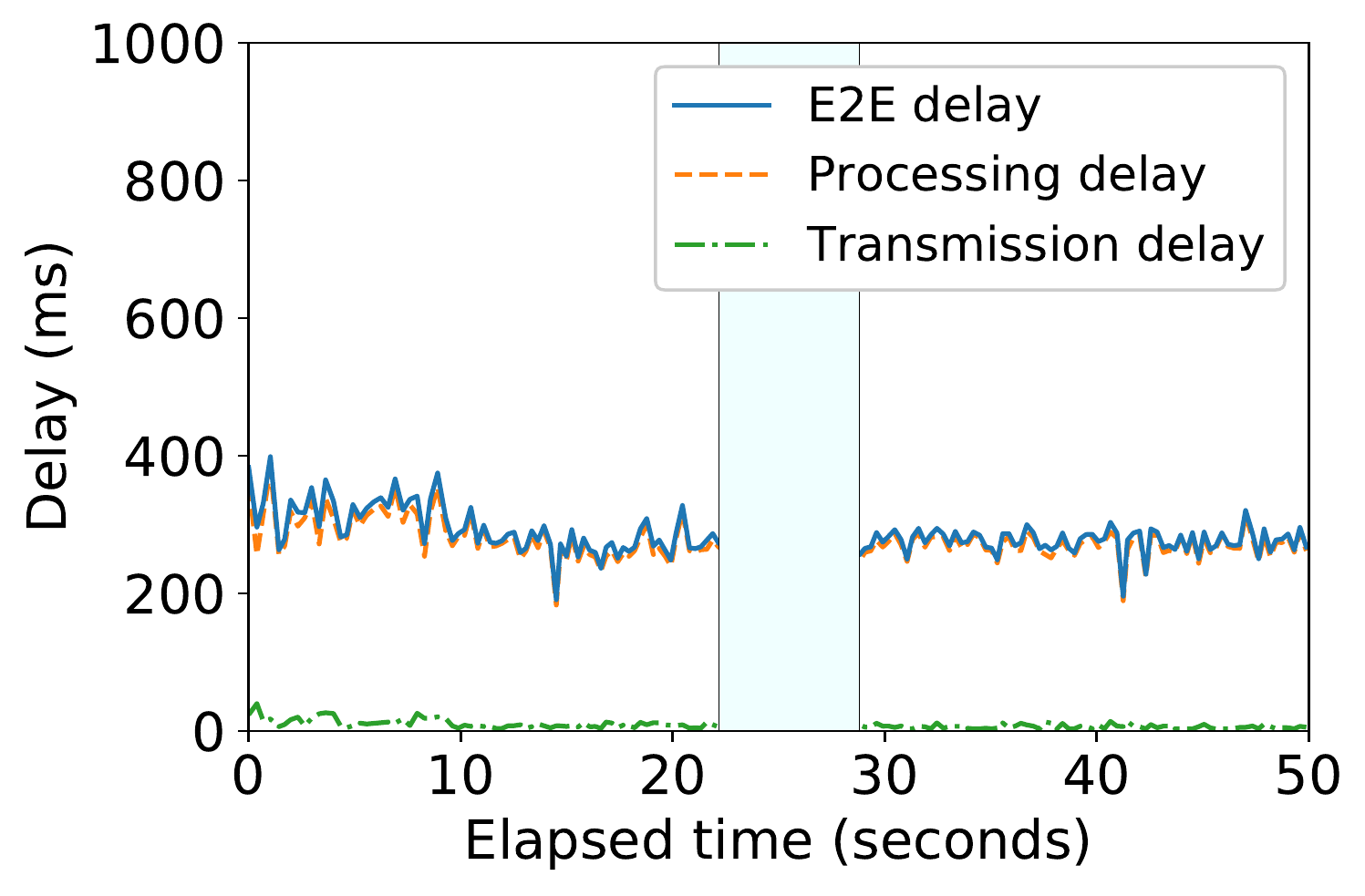}
		\label{fig:E2EdelayOpenfaceSimOptimization}%
	}\!\!\!%
	\caption{E2E delay of a MU who offloads tasks to \textit{Openface} service under four evaluated planners. The shaded gaps indicate service downtime caused by migration and handover.
         %In (b), (c), handover process and pre-migration phase are triggered at the beginning of the first gap, and the second gap is a migration period.
         In (b), (c), the first gap is handover period (which also triggers pre-migration phase in background), the second gap is migration period.
     }
	\label{fig:E2EdelayOpenfaceSim}
\end{figure*}

Second, we zoom in onto a 50\,s interval for a closer investigation of the E2E delay (including its two components) of a MU who offloads tasks to \textit{Openface} service under four planners in Fig.~\ref{fig:E2EdelayOpenfaceSim}. During this period, the MU moves from one BS to another BS.
%Among the planners, the E2E delay of the cloud planner is much higher than the other planners.
In Fig.~\ref{fig:E2EdelayOpenfaceSimCloud}, the cloud planner incurs much longer transmission delay despite its slightly shorter processing delay, resulting in a much higher E2E delay in comparison with the other planners.
%We see that the E2E delay with the cloud planner is much higher than that with the other planners.
%This is mainly due to the much longer transmission delay incurred by the cloud planner, despite its slightly shorter processing delay.

As shown in Fig. \ref{fig:E2EdelayOpenfaceSimRandom}, the random planner has a fluctuating E2E delay.
Although, before BS handover, the random planner may obtain low E2E delay by occasionally allocating the container to a nearby edge server, the E2E delay is significantly increased after BS handover because the next offloading server can be the old edge server or another server that is far away from the MU.
\iffalse
The E2E delay fluctuation also explains why the CDF lines of the random planner have a step format, as shown in Fig. \ref{fig:CDFServicesHomoNet}.
\fi

For the random planner and the nearest planner, we can see in Figs.~\ref{fig:E2EdelayOpenfaceSimRandom}, \ref{fig:E2EdelayOpenfaceSimNearest} that during the handover period, the E2E delay are significantly elevated.
%As shown in Figs. \ref{fig:E2EdelayOpenfaceSimRandom}, \ref{fig:E2EdelayOpenfaceSimNearest}, during the handover, the E2E delay under the random and nearest planners are significantly elevated.
This is mainly due to the much prolonged transmission delay after the MU handover to a new BS but the offloaded service is still running on the previous server (i.e., not migrated yet).
%In addition, the performance of the service is affected by the container migration process which consumes computational resource, resulting in increased processing delays.
%Consequently, the E2E delay adds up a high processing delay due to the pre-migration process that consumes CPU and affects to the performance of the container, and an increasing transmission delay that is affected by data transmission between two edge servers.

%The orchestration planner provides to the MU a smoother E2E delay than the other planners as shown in Fig.~\ref{fig:E2EdelayOpenfaceSimOptimization}.
As shown in Fig.~\ref{fig:E2EdelayOpenfaceSimOptimization}, the orchestrated planner has less fluctuation in E2E delay than the other planners, which implies a much smoother user experience.
This is because the orchestrated planner always allocates containers to the best server, and initiates container migration at the coordinated time with BS handover.
%So it not only provides low E2E delay but also minimizes service downtime.
%With the orchestrated migration-handover mechanism and placement optimization, the orchestration planner always allocates to the best server and triggers pre-migration, migration, and handover in an orchestrated manner to minimize service downtime.
Also because of the coordinated migration-handover mechanism, after the MU handover to a new BS, the server has also been migrated and hence the transmission delay is minimized.
Overall, the orchestrated planner achieves the lowest E2E delay among all the planners.

%----------------------------------------------
\subsubsection{Total service downtime}
\label{subsec:ResultDowntimeService}
%----------------------------------------------

Fig.~\ref{fig:E2EdelayOpenfaceSim} also shows the service downtime (indicated by the shaded gaps) experienced by the MU who offloads tasks to \textit{Openface} service under four planners.
The cloud planner has the shortest total service downtime because it only requires BS handover and not container migration.
However, it has significantly longer E2E delay as indicated in both Fig.~\ref{fig:E2EDelays} and Fig.~\ref{fig:E2EdelayOpenfaceSim}.
Now we zoom onto the service downtime of the other three planners which involves both container migration and BS handover.
We see in Fig.~\ref{fig:E2EdelayOpenfaceSim} that the random and nearest planners have a large total service downtime consisting of two separate periods, i.e., %a handover period (the first gap) and a migration (the second gap).
handover and migration periods.
Between the two periods, the offloaded service is still hosted on the old server (and under pre-migration phase), hence leads to a higher E2E delay as shown in Figs~\ref{fig:E2EDelayOpenface}, \ref{fig:E2EDelayYolo}.
In contrast, the orchestrated planner has a single and shorter downtime period, due to its orchestrated timing that takes into account both migration and handover.

Fig. \ref{fig:DowntimeCompared4planners} shows the total service downtime of a MU who offloads tasks to one of the three offloaded services under three migration planners over the whole experiment period. % of 1600s.
%Fig. \ref{fig:DowntimeCompared4planners}. compares the total service downtime (mean and standard deviation) of a MU running three offloaded services under the four planners.
% We can see that our orchestration planner reduces the service downtime of the random and nearest planners by more than 30\% for all the three services.
We can see that, in all three offloaded services, the orchestrated planner outperforms the random and nearest planners by reducing the total service downtime by 30-40\%.
%Although the cloud planner has the shortest service downtime, it suffers from high E2E delay (cf. Fig.~\ref{fig:E2EdelayOpenfaceSim}).
%Note that the impact of E2E delay on user experience is long-term and continuous whereas service downtime happens only once of each handover.
%Therefore, the orchestrated planner is still much more preferable than the cloud planner.
\iffalse % TODO: out of space
The 95\% confidence intervals of the total service downtime in Fig. \ref{fig:DowntimeCompared4planners} are wider than those of the E2E delay in Fig.~\ref{fig:E2EDelays}
because the number of events (around 6) that caused service downtime was much smaller than the number of measurements of E2E delay (more than 2500 measurements) during the experiments.
\fi

\begin{table}[ht]
	\caption{Checkpoint File Size Comparison}
	\label{tab:SpecificationService}
	\centering
    \footnotesize
	\begin{tabular}{ c@{\hspace{1em}}c@{\hspace{1em}} c@{\hspace{1em}}  c@{\hspace{1em}} c @{\hspace{1em}}}
		\toprule
		\textbf{Docker} & \textbf{Docker} &	\textbf{Checkpoint of} & \textbf{Checkpoint}  & \textbf{Reduction}\\
		\textbf{container} & \textbf{image size } & \textbf{pre-migration} &  \textbf{of migration} & \textbf{ratio} \\
		\midrule
		Simple   & 74.2 MB & 11.29 MB  & 47.7 KB   & \textbf{99.6}\% \\
		Openface & 1.86 GB & 196.8 MB  & 7.94 MB   & \textbf{96.0}\% \\
		Yolo     & 792 MB    & 584.8 MB  & 5.60 MB & \textbf{99.1}\% \\
		\bottomrule
	\end{tabular}
\end{table}
An important determining factor of the duration of migration, which contributes to the total service downtime, is the size of {\em checkpoint files} transferred during pre-migration and migration.
Therefore, we also present these details in Table \ref{tab:SpecificationService}.
As we can see, with our proposed delta checkpoint technique for container migration, the size of checkpoint files of migration is significantly smaller than that of the checkpoint files of pre-migration.
Specifically, the migration of Openface and Yolo services only transfer 7.94 MB and 5.6 MB which are just 4\% and 0.9\% of the pre-migration's checkpoint files.
This remarkably helps to reduce the migration time and hence minimize the total service downtime.
%\red{The reduction ratio is various between offloaded services due to their internal implementations and allocation of running memory.}

%In addition, Fig. 6 also shows that the total service downtime increases as the complexity of the offloaded task increases, Table \ref{tab:SpecificationService} provides the  quantitative reason.

In summary, the orchestrated planner with the delta checkpoint technique not only achieves the lowest E2E delay but also minimizes the total service downtime during BS handover.

% amount of data transmission for the migration which directly causes interrupt service.
%Since the pre-migration already brings most of the running memory to the destination server, the migration just takes care a small amount of changed memory state.

%----------------------------------------------
\section{Conclusion}
\label{sec:conclusions}
%----------------------------------------------

%This paper proposes a coordinated migration-handover mechanism that is enabled by a hierarchical MEC system architecture, in order to address the joint challenge of performing container migration and base station handover in a coordinated manner.
To address the joint challenge of performing container migration and base station handover, this paper proposes a coordinated migration-handover mechanism enabled by a hierarchical MEC system architecture.
We (1) formulate an optimization problem for container placement and base station allocation, and (2) derive the best time to trigger handover, pre-migration, and migration, based on a delta checkpoint technique that we propose.
We then set up a real MEC testbed, and implement our proposed mechanism in an orchestrated planner as well as three other baseline planners for comparison.
%We also evaluate these planners' performance using real-world applications on a real MEC testbed that we set up.
The experimental results demonstrate that our proposed mechanism outperforms other solutions by significantly reducing E2E delay and service downtime for mobile users.
Our work contributes toward offering much smoother user experience in MEC, especially for time-sensitive applications.
%can be consequential in offering smooth user experience

%\end{document}  % This is where a 'short' article might terminate

% ------------------------------------------------------%
%\section*{Acknowledgement}
% ------------------------------------------------------%

% ------------------------------------------------------ %

% ------------------------------------------------------%

\bibliographystyle{IEEEtran}
%\scriptsize{
\bibliography{references}
%}

% ------------------------------------------------------%

%\appendix
%Appendix A
%\section{Headings in Appendices}

%\begin{acks}
%  The authors would like to thank XXX for providing the
%  xxx code of the xxx.

  %The authors would also like to thank the anonymous referees for
  %their valuable comments and helpful suggestions. The work is
  %supported by the \grantsponsor{GS501100001809}{National Natural
  %  Science Foundation of
  %  China}{http://dx.doi.org/10.13039/501100001809} under Grant
  %No.:~\grantnum{GS501100001809}{61273304}
  %and~\grantnum[http://www.nnsf.cn/youngscientists]{GS501100001809}{Young
  %  Scientists' Support Program}.

%\end{acks}

\end{document}